\documentclass[pra]{revtex4}
 \usepackage{amssymb} \usepackage{graphicx}

\begin{document}
 \title{Symmetry operators of Killing spinors and superalgebras in $AdS_5$}

\author{\"Umit Ertem}
 \email{umitertemm@gmail.com}
\address{School of Mathematics, The University of Edinburgh, James Clerk Maxwell Building, Peter Guthrie Tait Road, Edinburgh, EH9 3FD, United Kingdom\\}

\begin{abstract}

We construct the first-order symmetry operators of Killing spinor equation in terms of odd Killing-Yano forms. By modifying the Schouten-Nijenhuis bracket of Killing-Yano forms, we show that the symmetry operators of Killing spinors close into an algebra in $AdS_5$ spacetime. Since the symmetry operator algebra of Killing spinors corresponds to a Jacobi identity in extended Killing superalgebras, we investigate the possible extensions of Killing superalgebras to include higher-degree Killing-Yano forms. We found that there is a superalgebra extension but no Lie superalgebra extension of the Killing superalgebra constructed out of Killing spinors and odd Killing-Yano forms in $AdS_5$ background.

\end{abstract}

\maketitle

\section{Introduction}

Constructing symmetry operators of an equation is an important step to obtain its solutions. The commuting symmetry operators can give way to find the solutions of the equation by the method of separation of variables \cite{Miller}. First-order symmetry operators of the Dirac equation in curved backgrounds can be constructed from Killing-Yano (KY) forms which correspond to the antisymmetric generalizations of Killing vector fields \cite{Benn Charlton, Benn Kress, Acik Ertem Onder Vercin1}. Although in some special cases, which constrain the properties of KY forms, they can satisfy a closed symmetry algebra \cite{Cariglia Krtous Kubiznak}, this is not the case in general and one has to construct higher order operators. A subset of the solutions of the Dirac equation is Killing spinors which are solutions of the twistor equation at the same time \cite{Lichnerowicz, Baum}. Killing spinors play an important role in supergravity theories since the supergravity Killing spinors, which are parallel spinors with respect to a supergravity spin connection and correspond to the preserved supersymmetries of the supergravity background, can be constructed from the Killing spinors by the cone construction \cite{Bar}. Hence, finding the solutions of the Killing spinor equation  gives way to determine the supergravity Killing spinors in relevant supergravity backgrounds. One way to find the solutions of Killing spinor equation is constructing the symmetry operators and since they are a subset of the solutions of the Dirac equation, the symmetry operators of both equations can be related to each other.

Killing spinors also play a role in the construction of the Killing superalgebras which determine the geometrical peoperties of supergravity backgrounds \cite{OFarrill Meessen Philip, OFarrill HackettJones Moutsopoulos}. They constitute the odd part of the superalgebra and the squaring map of Killing spinors generates Killing vector fields which correspond to the even part of it. However, the squaring map also generates higher-degree forms and they correspond to KY forms for the Killing spinor case \cite{Acik Ertem}. So, the possibilities of the extensions of Killing superalgebras to include higher-degree KY forms is worth to investigate. On the other hand, Killing spinors generate only a subset of KY forms through the squaring map in general and there may be KY forms that cannot be obtained from Killing spinors while constructing symmetry operators of Killing spinors from KY forms can give way to find all the Killing spinors of the background. The classification of the manifolds that have parallel spinors and Killing spinors has been done by using G-structures \cite{Bar, OFarrill}. It is known that the manifolds that have Killing spinors also have KY forms \cite{Papadopoulos}. However, there are also manifolds that have KY forms and are not in the classification table of the manifolds that have Killing spinors. One example for this case is the nearly K-cosymplectic manifolds for which the fundamental form corresponds to a KY 2-form \cite{Santillan}.

In this paper, we show that the first-order symmetry operators of the Killing spinor equation can be induced from the first-order symmetry operators of the Dirac equation by restricting them to the odd KY forms. KY forms satisfy a graded Lie algebra structure under Schouten-Nijenhuis (SN) bracket in constant curvature spacetimes \cite{Kastor Ray Traschen, Ertem Acik}. However, the first-order symmetry operators do not satisfy a closed algebra under this bracket. To obtain a symmetry algebra of Killing spinors, we concentrate on five dimensional constant curvature spacetimes and define a modified KY bracket. Although we choose the five dimensional anti-de Sitter ($AdS_5$) spacetime because of the importance of it in string theory and supergravity backgrounds, the analysis is also relevant in other five and lower dimensional constant curvature backgrounds. We show that the modified KY bracket gives again a KY form and prove that the first-order symmetry operators of Killing spinors satisfy a symmetry algebra with respect to this bracket. Since the symmetry algebra property of Killing spinors correspond to one of the Jacobi identities of extended Killing superalgebra, we investigate the conditions on the extension of Killing superalgebras in $AdS_5$. We found that there is no extension of the Killing superalgebra since the Jacobi identities constrain the possibilities of Lie brackets between superalgebra elements, but the Killing spinors and odd KY forms constitute a superalgebra since the bilinear maps of the superalgebra can be defined, although it is not a Lie superalgebra.

The paper is organized as follows. In Section 2, we construct the first-order symmetry operators of Killing spinor equation from odd degree KY forms. In a subsection, we demonstrate the graded Lie algebra structure of KY forms under SN bracket and by concentrating on $AdS_5$ spacetime, we define a new modified KY bracket. Another subsection includes the proof of the first-order symmetry operator algebra of Killing spinors with respect to the modified KY bracket. In Section 3, we consider the construction of Killing superalgebras and investigate the conditions for extending them to include higher-degree KY forms and symmetry operators of Killing spinors. Section 4 concludes the paper.

\section{First-Order Symmetry Operator Algebra of Killing Spinors}

In a spin manifold $M$, spinor fields are defined as the sections of the spin bundle $S(M)$ which can be induced from the Clifford bundle $Cl(M)$. We consider two first-order differential operators defined on spinor fields. The first one is the Dirac operator defined as follows
\begin{equation}
\displaystyle{\not}D=e^a.\nabla_{X_a}
\end{equation}
where $\{X_a\}$ is orthonormal frame basis and $\{e^a\}$ is orthonormal co-frame basis that satisfy $e^a(X_b)=\delta^a_b$. $\nabla$ is the spinor covariant derivative defined on spinor fields and $.$ denotes the Clifford product. Spinor fields that are in the kernel of the Dirac operator are called harmonic spinors and correspond to the solutions of the massless Dirac equation. The massive Dirac equation for a spinor $\psi$ is written as follows
\begin{equation}
\displaystyle{\not}D\psi=m\psi
\end{equation}
where $m$ denotes the mass and the solutions of the Dirac equation are eigenspinors of the Dirac operator. The second first-order differential operator that we consider is the twistor or Penrose operator defined as
\begin{equation}
{\cal{P}}_X=\nabla_X-\frac{1}{n}\widetilde{X}.\displaystyle{\not}D
\end{equation}
where $X$ is any vector field with $\widetilde{X}$ its metric dual and $n$ is the dimension of the manifold. The spinor fields that are in the kernel of the twistor operator are called twistor spinors and satisfy the following twistor equation
\begin{equation}
\nabla_X\psi=\frac{1}{n}\widetilde{X}.\displaystyle{\not}D\psi.
\end{equation}
Twistor spinors that are also eigenspinors of the Dirac operator at the same time are called Killing spinors. They are solutions of the Killing spinor equation which is written as follows
\begin{equation}
\nabla_X\psi=\lambda\widetilde{X}.\psi
\end{equation}
where $\lambda$ is a complex constant called Killing number which is real or pure imaginary. The existence of Killing spinors constrains the geometry such that the scalar curvature of the manifold is equal to
\begin{equation}
{\cal{R}}=-4\lambda^2n(n-1).
\end{equation}
A special class of manifolds that satisfy this condition are constant curvature spacetimes with curvature $R_{ab}=-4\lambda^2e_a\wedge e_b$. For $\lambda$ real these correspond to negative curvature manifolds such as anti-de Sitter ($AdS$) spacetimes and for $\lambda$ pure imaginary they correspond to positive curvature manifolds such as de Sitter ($dS$) spacetimes.

Symmetry operators play important roles in finding solutions of differential equations. Solutions of Dirac equation (2) are preserved under the Lie derivative operation on spinors with respect to a Killing vector field, namely the Lie derivative of a solution is again a solution. This means that the Lie derivative with respect to a Killing vector is a symmetry operator of the Dirac equation. For massles Dirac equation this is also the case for the Lie derivative with respect to conformal Killing vectors. The Lie derivative with respect to a Killing vector field $K$ is defined for an inhomogeneous differential form $\alpha$ as follows \cite{Benn Tucker}
\begin{equation}
{\cal{L}}_K\alpha=\nabla_K\alpha+\frac{1}{4}[d\widetilde{K}, \alpha]_{Cl}
\end{equation}
where $[. , .]_{Cl}$ is the Clifford commutator. The Lie derivative on spinor fields with respect to a Killing vector field can be defined from (7) by projecting the second term on the minimal left ideal of the Clifford algebra \cite{Kosmann};
\begin{equation}
{\cal{L}}_K\psi=\nabla_K\psi+\frac{1}{4}d\widetilde{K}.\psi.
\end{equation}
Other first-order symmetry operators of the Dirac equation can be constructed from higher degree forms. These higher degree forms are antisymmetric generalizations of Killing vector fields and are called Killing-Yano (KY) forms \cite{Yano}. A KY $p$-form $\omega$ satisfy the following equation
\begin{equation}
\nabla_X\omega=\frac{1}{p+1}i_Xd\omega
\end{equation}
where $X$ is an arbirtary vector field, $i_X$ is the interior derivative operator and $d$ is the exterior derivative. First-order symmetry operators of Dirac equation constructed out of KY $p$-forms $\omega$ are written as follows \cite{Benn Charlton, Benn Kress, Acik Ertem Onder Vercin1}
\begin{equation}
L_{\omega}\psi=(i_{X^a}\omega).\nabla_{X_a}\psi+\frac{p}{2(p+1)}d\omega.\psi.
\end{equation}
For the special case of $p=1$, this symmetry operator reduces to the Lie derivative of a spinor with respect to a Killing vector field in (8). In this sense, the symmetry operators of Dirac equation can be seen as generalized Lie derivatives of spinors with respect to KY forms and the action of these operators to inhomogeneous differential forms $\alpha$ can be written as
\begin{equation}
L_{\omega}\alpha=(i_{X^a}\omega).\nabla_{X_a}\alpha+\frac{p}{2(p+1)}[d\omega, \alpha ]_{Cl}.
\end{equation}
On the other hand, the symmetry operators of massless Dirac equation can be constructed out of conformal KY forms which are antisymmetric generalizations of conformal Killing vectors. Similarly, the first-order operators taking solutions of the twistor equation to the solutions of the Dirac equation can also be constructed from conformal KY forms in conformally flat spacetimes \cite{Benn Kress2}.

Although the operator in (10) is a symmetry operator for the Dirac equation, this does not mean that it can also preserve the subset of Killing spinors. However, we will see that this can be possible for some special cases. By using the definition of Killing spinor equation in (5), the action of (10) on a Killing spinor $\epsilon$ reads
\begin{eqnarray}
L_{\omega}\epsilon&=&(i_{X^a}\omega).\nabla_{X_a}\epsilon+\frac{p}{2(p+1)}d\omega.\epsilon\nonumber\\
&=&\lambda(i_{X^a}\omega).e_a.\epsilon+\frac{p}{2(p+1)}d\omega.\epsilon\nonumber\\
&=&(-1)^{p-1}\lambda p\omega.\epsilon+\frac{p}{2(p+1)}d\omega.\epsilon
\end{eqnarray}
where we have used the equality $i_{X^a}\omega.e_a=(-1)^{p-1}e_a\wedge i_{X^a}\omega=(-1)^{p-1}p\omega$. In fact, the symmetry operators in (10) constructed out of odd degree KY forms also preserve the solutions of the Killling spinor equation in constant curvature spacetimes, namely if $\epsilon$ is a Killing spinor, then $L_{\omega}\epsilon$ is also a Killing spinor in that case. So, we have the equality $\nabla_XL_{\omega}\epsilon=\lambda\widetilde{X}.L_{\omega}\epsilon$. This can be proven as follows. The left hand side can be calculated from (12) for a frame basis $X_a$ and for an odd KY $p$-form $\omega$ as
\begin{eqnarray}
\nabla_{X_a}L_{\omega}\epsilon&=&\nabla_{X_a}\left(\lambda p\omega.\epsilon+\frac{p}{2(p+1)}d\omega.\epsilon\right)\nonumber\\
&=&\lambda p\nabla_{X_a}\omega.\epsilon+\lambda p\omega.\nabla_{X_a}\epsilon+\frac{p}{2(p+1)}\nabla_{X_a}d\omega.\epsilon+\frac{p}{2(p+1)}d\omega.\nabla_{X_a}\epsilon\nonumber\\
&=&\lambda\frac{p}{p+1}i_{X_a}d\omega.\epsilon+\lambda^2p\omega.e_a.\epsilon+\frac{1}{2}(R_{ba}\wedge i_{X_b}\omega).\epsilon+\lambda\frac{p}{2(p+1)}d\omega.e_a.\epsilon
\end{eqnarray}
where we have used (9), (5) and the integrability condition of KY forms which is
\begin{equation}
\nabla_{X_a}d\omega=\frac{p+1}{p}R_{ba}\wedge i_{X^b}\omega
\end{equation}
(see \cite{Acik Ertem Onder Vercin2} for a proof). The Clifford product of a 1-form $\widetilde{X}$ with a $p$-form $\alpha$ can be written in terms of wedge product and interior derivative as follows
\begin{eqnarray}
\widetilde{X}.\alpha&=&\widetilde{X}\wedge\alpha+i_{X}\alpha\\
\alpha.\widetilde{X}&=&\widetilde{X}\wedge\eta\alpha-i_{X}\eta\alpha
\end{eqnarray}
where $\eta$ acts on $p$-forms as $\eta\alpha=(-1)^p\alpha$. By using these equalities and the curvature 2-forms $R_{ba}=-4\lambda^2e_b\wedge e_a$, we obtain
\begin{eqnarray}
\nabla_{X_a}L_{\omega}\epsilon&=&\lambda\frac{p}{p+1}i_{X_a}d\omega.\epsilon+\lambda^2p(-e_a\wedge\omega+i_{X_a}\omega).\epsilon-2\lambda^2(e_b\wedge e_a\wedge i_{X^b}\omega).\epsilon\nonumber\\
&&+\lambda\frac{p}{2(p+1)}(e_a\wedge d\omega-i_{X_a}d\omega).\epsilon\nonumber\\
&=&\lambda^2p(e_a\wedge\omega).\epsilon+\lambda^2pi_{X_a}\omega.\epsilon+\lambda\frac{p}{2(p+1)}(e_a\wedge d\omega).\epsilon+\lambda\frac{p}{2(p+1)}i_{X_a}d\omega.\epsilon
\end{eqnarray}
where we have used $e_b\wedge i_{X^b}\omega=p\omega$. On the other hand, we can obtain the following equality for the right hand side of the Killing spinor equation of $L_{\omega}\epsilon$
\begin{eqnarray}
\lambda e_a.L_{\omega}\epsilon&=&\lambda^2pe_a.\omega.\epsilon+\lambda\frac{p}{2(p+1)}e_a.d\omega.\epsilon\nonumber\\
&=&\lambda^2p(e_a\wedge\omega+i_{X_a}\omega).\epsilon+\lambda\frac{p}{2(p+1)}(e_a\wedge d\omega+i_{X_a}d\omega).\epsilon.
\end{eqnarray}
By comparing (17) and (18), one can see that
\begin{equation}
\nabla_{X_a}L_{\omega}\epsilon=\lambda e_a.L_{\omega}\epsilon.
\end{equation}
This shows that the symmetry operators in (10) constructed out of odd KY forms preserve the solutions of the Killing spinor equation in constant curvature spacetimes. In fact, they also correspond to symmetry operators of Killing spinors in more general manifolds, but this time they must be constructed out of odd special KY forms. Special KY forms satisfy the following condition \cite{Semmelmann, Lischewski}
\begin{eqnarray}
\nabla_Xd\omega=-\frac{p+1}{n(n-1)}{\cal{R}}\widetilde{X}\wedge\omega\nonumber.
\end{eqnarray}
This is the special case of the integrability condition in (14) and in constant curvature spacetimes every KY form satisfies this condition, so all KY forms are special KY forms in that case. As can be seen from a similiar analysis given in (12) to (19), the even KY forms does not generate a symmetry operator for the Killing spinor equation. The natural question to ask at this point is that whether the first-order symmetry operators constructed for Killing spinor equation can close into an algebra structure as in the case of Killing vector fields that satisfy $[{\cal{L}}_X, {\cal{L}}_Y]={\cal{L}}_{[X, Y]}$. To answer this question, one needs to find a relevant bracket for KY forms and this will be the task in the next subsection.

\subsection{A New Bracket for KY Forms}

The generalization of the Lie bracket of vector fields to higher degree forms corresponds to Schouten-Nijenhuis (SN) bracket \cite{Schouten, Nijenhuis}. This bracket is defined for a $p$-form $\alpha$ and a $q$-form $\beta$ as follows
\begin{equation}
[\alpha,\beta]_{SN}=i_{X^a}\alpha\wedge\nabla_{X_a}\beta+(-1)^{pq}i_{X^a}\beta\wedge\nabla_{X_a}\alpha
\end{equation}
which gives a $(p+q-1)$-form and reduces to the metric dual of the ordinary Lie bracket of vector fields for $p=q=1$. SN bracket satisfies the following graded Lie bracket properties
\begin{equation}
[\alpha, \beta]_{SN}=(-1)^{pq}[\beta, \alpha]_{SN}
\end{equation}
\begin{equation}
(-1)^{p(r+1)}[\alpha, [\beta, \gamma]_{SN}]_{SN}+(-1)^{q(p+1)}[\beta, [\gamma, \alpha]_{SN}]_{SN}+(-1)^{r(q+1)}[\gamma, [\alpha, \beta]_{SN}]_{SN}=0
\end{equation}
where $\gamma$ is an $r$-form. While Killing vector fields constitute a Lie algebra with respect to the Lie bracket of vector fields in all cases, KY forms satisfy a graded Lie algebra or a Lie superalgebra structure in constant curvature spacetimes \cite{Kastor Ray Traschen, Ertem Acik}. To prove this, let us consider a KY $p$-form $\omega_1$ and a KY $q$-form $\omega_2$ and compute the covariant derivative of their SN bracket
\begin{eqnarray}
\nabla_{X_b}[\omega_1, \omega_2]_{SN}&=&\nabla_{X_b}\big(i_{X^a}\omega_1\wedge\nabla_{X_a}\omega_2+(-1)^{pq}i_{X^a}\omega_2\wedge\nabla_{X_a}\omega_1\big)\nonumber\\
&=&\nabla_{X_b}i_{X^a}\omega_1\wedge\nabla_{X_a}\omega_2+i_{X^a}\omega_1\wedge\nabla_{X_b}\nabla_{X_a}\omega_2\nonumber\\
&&+(-1)^{pq}\nabla_{X_b}i_{X^a}\omega_2\wedge\nabla_{X_a}\omega_1+(-1)^{pq}i_{X^a}\omega_2\wedge\nabla_{X_b}\nabla_{X_a}\omega_1\nonumber\\
&=&\frac{1}{(p+1)(q+1)}\left(i_{X^a}i_{X_b}d\omega_1\wedge i_{X_a}d\omega_2+(-1)^{pq}i_{X^a}i_{X_b}d\omega_2\wedge i_{X_a}d\omega_1\right)\nonumber\\
&&+\frac{1}{q}i_{X^a}\omega_1\wedge i_{X_a}(R_{cb}\wedge i_{X^c}\omega_2)+\frac{(-1)^{pq}}{p}i_{X^a}\omega_2\wedge i_{X_a}(R_{cb}\wedge i_{X^c}\omega_1)
\end{eqnarray}
where we have used (9), the definition of the curvature operator $[\nabla_{X_a}, \nabla_{X_b}]=R(X_a, X_b)+\nabla_{[X_a, X_b]}$ and (14). On the other hand, if we apply first $d$ and then $i_{X_b}$ to the SN bracket of KY forms, we obtain
\begin{eqnarray}
i_{X_b}d[\omega_1, \omega_2]_{SN}&=&i_{X_b}di_{X^a}\omega_1\wedge\nabla_{X_a}\omega_2+(-1)^pdi_{X^a}\omega_1\wedge i_{X_b}\nabla_{X_a}\omega_2\nonumber\\
&&-(-1)^pi_{X_b}i_{X^a}\omega_1\wedge d\nabla_{X_a}\omega_2+i_{X^a}\omega_1\wedge i_{X_b}d\nabla_{X_a}\omega_2\nonumber\\
&&+(-1)^{pq}i_{X_b}di_{X^a}\omega_2\wedge\nabla_{X_a}\omega_1+(-1)^{p(q+1)}di_{X^a}\omega_2\wedge i_{X_b}\nabla_{X_a}\omega_1\nonumber\\
&&-(-1)^{p(q+1)}i_{X_b}i_{X^a}\omega_2\wedge d\nabla_{X_a}\omega_1+(-1)^{pq}i_{X^a}\omega_2\wedge i_{X_b}d\nabla_{X_a}\omega_1\nonumber\\
&=&\frac{p+q}{(p+1)(q+1)}\left(i_{X^a}i_{X_b}d\omega_1\wedge i_{X_a}d\omega_2+(-1)^{pq}i_{X_a}i_{X_b}d\omega_2\wedge i_{X^a}d\omega_1\right)\nonumber\\
&&-(-1)^{p}\left(\frac{1}{p}+\frac{1}{q}\right)i_{X_b}\left(i_{X^a}\omega_1\wedge R_{ca}\wedge i_{X^c}\omega_2\right)
\end{eqnarray}
where we have used (9), the definition of the curvature operator and $[\nabla_X, d]=0$. By comparing (23) and (24), one can see that the SN bracket of KY form does not satisfy the KY equation in general. However, for constant curvature spacetimes $R_{ab}=ce_a\wedge e_b$, the curvature terms of both equations cancel each other and one obtains the KY equation
\begin{equation}
\nabla_{X_b}[\omega_1, \omega_2]_{SN}=\frac{1}{p+q}i_{X_b}d[\omega_1, \omega_2]_{SN}.
\end{equation}
This shows the graded Lie algebra structure of KY forms in constant curvature spacetimes. Indeed, it corresponds to a Lie superalgebra $\mathfrak{k}=\mathfrak{k}_0\oplus\mathfrak{k}_1$ with the Lie algebra of odd KY forms $\mathfrak{k}_0$ and the space of even KY forms $\mathfrak{k}_1$. So, the odd KY forms satisfy a Lie algebra under SN bracket and the algebra structure of symmetry operators of Killing spinor equation in terms of odd KY forms can be investigated.

Although the SN bracket is a generalization of the Lie bracket of Killing vector fields to higher degree KY forms, the first-order symmetry operators of the Killing spinor equation do not satisfy an algebra with respect to SN bracket in general, namely $[L_{\omega_1}, L_{\omega_2}]\neq L_{[\omega_1, \omega_2]_{SN}}$. However, we will consider the five dimensional constant curvature spacetimes, especially $AdS_5$, and define a new bracket of KY forms to obtain an algebra structure of first-order symmetry operators of the Killing spinor equation. The results will be relevant for other five dimensional constant curvature spacetimes such as $dS_5$ and Minkowski spacetimes and also for lower dimensions. We choose $AdS_5$ as an example, because of the importance of it in string theory and supergravity backgrounds.

The new modified bracket of KY forms in $AdS_5$ that we propose is
\begin{equation}
[\omega_1, \omega_2]_{KY}=\frac{pq}{p+q-1}[\omega_1, \omega_2]_{SN}-\frac{pq}{p+q}[i_{X_a}i_{X_b}\omega_1, i_{X^a}i_{X^b}\omega_2]_{SN}.
\end{equation}
In five dimensions, odd KY forms are 1-forms, 3-forms and 5-form. KY 5-form corresponds to the volume form with a constant coefficient. So, one can see from the definition of the SN bracket that the second term on the right hand side of (26) is non-zero only for $p=q=3$. We know that the first term on the right hand side of (26) gives a KY form, but we have to prove that this is also the case for the second term. It is enough to show this for $p=q=3$ and if we calculate the covariant derivative of the second term on the right hand side of (26), we find
\begin{eqnarray}
\nabla_{X_d}[i_{X_b}i_{X_c}\omega_1, i_{X^b}i_{X^c}\omega_2]_{SN}&=&\frac{1}{q+1}\nabla_{X_d}(i_{X_a}i_{X_b}i_{X_c}\omega_1)i_{X^a}i_{X^b}i_{X^c}d\omega_2\nonumber\\
&&+\frac{1}{q+1}(i_{X_a}i_{X_b}i_{X_c}\omega_1)\nabla_{X_d}i_{X^a}i_{X^b}i_{X^c}d\omega_2\nonumber\\
&&-\frac{1}{p+1}\nabla_{X_d}(i_{X^a}i_{X^b}i_{X^c}\omega_2)i_{X_a}i_{X_b}i_{X_c}d\omega_1\nonumber\\
&&-\frac{1}{p+1}(i_{X^a}i_{X^b}i_{X^c}\omega_2)\nabla_{X_d}i_{X_a}i_{X_b}i_{X_c}d\omega_1\nonumber\\
&=&\frac{1}{(p+1)(q+1)}(i_{X_a}i_{X_b}i_{X_c}i_{X_d}d\omega_1)i_{X^a}i_{X^b}i_{X^c}d\omega_2\nonumber\\
&&+\frac{1}{q}(i_{X_a}i_{X_b}i_{X_c}\omega_1)i_{X^a}i_{X^b}i_{X^c}(R_{fd}\wedge i_{X^f}\omega_2)\nonumber\\
&&-\frac{1}{(p+1)(q+1)}(i_{X^a}i_{X^b}i_{X^c}i_{X_d}d\omega_2)i_{X_a}i_{X_b}i_{X_c}d\omega_1\nonumber\\
&&-\frac{1}{p}(i_{X^a}i_{X^b}i_{X^c}\omega_2)i_{X_a}i_{X_b}i_{X_c}(R_{fd}\wedge i_{X^f}\omega_1)
\end{eqnarray}
where we have used (9), (14) and $[\nabla_{X_a}, i_{X_b}]=i_{\nabla_{X_a}X_b}$. The existence of Killing spinors in $AdS_5$ imposes the condition that the curvature 2-forms are equal to $R_{ab}=-4\lambda^2e_a\wedge e_b$ (for $\lambda$ real). By using this, we obtain
\begin{eqnarray}
\nabla_{X_d}[i_{X_b}i_{X_c}\omega_1, i_{X^b}i_{X^c}\omega_2]_{SN}&=&\frac{1}{(p+1)(q+1)}\bigg((i_{X_a}i_{X_b}i_{X_c}i_{X_d}d\omega_1)i_{X^a}i_{X^b}i_{X^c}d\omega_2\nonumber\\
&&-(i_{X^a}i_{X^b}i_{X^c}i_{X_d}d\omega_2)i_{X_a}i_{X_b}i_{X_c}d\omega_1\bigg)\nonumber\\
&&+12\lambda^2\big((i_{X_a}i_{X_b}i_{X_d}\omega_1)i_{X^a}i_{X^b}\omega_2\nonumber\\
&&-(i_{X^a}i_{X^b}i_{X_d}\omega_2)i_{X_a}i_{X_b}\omega_1\big).
\end{eqnarray}
On the other hand, the action of $i_{X_d}d$ on the same bracket gives
\begin{eqnarray}
i_{X_d}d[i_{X_b}i_{X_c}\omega_1, i_{X^b}i_{X^c}\omega_2]_{SN}&=&\frac{1}{q+1}i_{X_d}\big(d(i_{X_a}i_{X_b}i_{X_c}\omega_1)\wedge i_{X^a}i_{X^b}i_{X^c}d\omega_2\big)\nonumber\\
&&+\frac{1}{q+1}i_{X_d}\big((i_{X_a}i_{X_b}i_{X_c}\omega_1)di_{X^a}i_{X^b}i_{X^c}d\omega_2\big)\nonumber\\
&&-\frac{1}{p+1}i_{X_d}\big(di_{X^a}i_{X^b}i_{X^c}\omega_2\wedge i_{X_a}i_{X_b}i_{X_c}d\omega_1\big)\nonumber\\
&&-\frac{1}{p+1}i_{X_d}\big((i_{X^a}i_{X^b}i_{X^c}\omega_2)di_{X_a}i_{X_b}i_{X_c}d\omega_1\big)\nonumber\\
&=&\left(\frac{p+q-4}{(p+1)(q+1)}\right)\bigg((i_{X_a}i_{X_b}i_{X_c}i_{X_d}d\omega_1)i_{X^a}i_{X^b}i_{X^c}d\omega_2\nonumber\\
&&-(i_{X^a}i_{X^b}i_{X^c}i_{X_d}d\omega_2)i_{X_a}i_{X_b}i_{X_c}d\omega_1\bigg)\nonumber\\
&&+24\lambda^2\big((i_{X_a}i_{X_b}i_{X_d}\omega_1)i_{X^a}i_{X^b}\omega_2\nonumber\\
&&-(i_{X^a}i_{X^b}i_{X_d}\omega_2)i_{X_a}i_{X_b}\omega_1\big)
\end{eqnarray}
where we have used ${\cal{L}}_{X_a}=i_{X_a}d+di_{X_a}$ which reduces to $\nabla_{X_a}=i_{X_a}d+di_{X_a}$ in normal coordinates. For $p=q=3$, the second term on the right hand side of (26) gives a 1-form and we can see from (28) and (29) that
\begin{equation}
\nabla_{X_d}[i_{X_b}i_{X_c}\omega_1, i_{X^b}i_{X^c}\omega_2]_{SN}=\frac{1}{2}i_{X_d}d[i_{X_b}i_{X_c}\omega_1, i_{X^b}i_{X^c}\omega_2]_{SN}
\end{equation}
which is the KY equation and this proves that the new bracket in (26) gives a KY form.

\subsection{Symmetry Operator Algebra in $AdS_5$}

After defining the new bracket for KY forms as in (26), we can search for the algebra structure of first-order symmetry operators of Killing spinor equation. For a KY $p$-form $\omega_1$ and a KY $q$-form $\omega_2$, where $p$ and $q$ are odd, one can write the action of two first-order symmetry operators on a Killing spinor $\epsilon$ by using (12) as
\begin{eqnarray}
L_{\omega_1}L_{\omega_2}\epsilon&=&L_{\omega_1}\left(\lambda q\omega_2.\epsilon+\frac{q}{2(q+1)}d\omega_2.\epsilon\right)\nonumber\\
&=&\lambda p\omega_1.\left(\lambda q\omega_2.\epsilon+\frac{q}{2(q+1)}d\omega_2.\epsilon\right)+\frac{p}{2(p+1)}d\omega_1.\left(\lambda q\omega_2.\epsilon+\frac{q}{2(q+1)}d\omega_2.\epsilon\right)\nonumber\\
&=&\lambda^2pq\omega_1.\omega_2.\epsilon+\lambda\frac{pq}{2(q+1)}\omega_1.d\omega_2.\epsilon\nonumber\\
&&+\lambda\frac{pq}{2(p+1)}d\omega_1.\omega_2.\epsilon+\frac{pq}{4(p+1)(q+1)}d\omega_1.d\omega_2.\epsilon
\end{eqnarray}
and a similar calculation gives
\begin{eqnarray}
L_{\omega_2}L_{\omega_1}\epsilon&=&\lambda^2pq\omega_2.\omega_1.\epsilon+\lambda\frac{pq}{2(p+1)}\omega_2.d\omega_1.\epsilon\nonumber\\
&&+\lambda\frac{pq}{2(q+1)}d\omega_2.\omega_1.\epsilon+\frac{pq}{4(p+1)(q+1)}d\omega_2.d\omega_1.\epsilon.
\end{eqnarray}
The difference of (31) and (32) corresponds the commutator of symmetry operators as follows
\begin{eqnarray}
[L_{\omega_1}, L_{\omega_2}]\epsilon&=&\lambda^2pq[\omega_1, \omega_2]_{Cl}.\epsilon+\lambda\frac{pq}{2(q+1)}[\omega_1, d\omega_2]_{Cl}.\epsilon\nonumber\\
&&+\lambda\frac{pq}{2(p+1)}[d\omega_1, \omega_2]_{Cl}.\epsilon+\frac{pq}{4(p+1)(q+1)}[d\omega_1, d\omega_2]_{Cl}.\epsilon.
\end{eqnarray}
On the other hand, the first-order symmetry operator constructed out of the new bracket of KY forms reads as
\begin{eqnarray}
L_{[\omega_1, \omega_2]_{KY}}\epsilon&=&\lambda\pi[\omega_1, \omega_2]_{KY}.\epsilon+\frac{\pi}{2(\pi+1)}d[\omega_1, \omega_2]_{KY}.\epsilon\nonumber\\
&=&\lambda pq[\omega_1, \omega_2]_{SN}.\epsilon-\lambda\frac{pq(p+q-5)}{(p+q)}[i_{X_a}i_{X_b}\omega_1, i_{X^a}i_{X^b}\omega_2]_{SN}.\epsilon\nonumber\\
&&+\frac{pq}{2(p+q)}d[\omega_1, \omega_2]_{SN}.\epsilon-\frac{pq(p+q-5)}{2(p+q)(p+q-4)}d[i_{X_a}i_{X_b}\omega_1, i_{X^a}i_{X^b}\omega_2]_{SN}.\epsilon
\end{eqnarray}
where we have used (26) and $\pi$ denotes the degree of the form that appears in the definition of the symmetry operator, for example, $\pi\left([\omega_1,\omega_2]_{SN}\right)=(p+q-1)[\omega_1,\omega_2]_{SN}$ and $\pi\left([i_{X_a}i_{X_b}\omega_1, i_{X^a}i_{X^b}\omega_2]_{SN}\right)=(p+q-5)[i_{X_a}i_{X_b}\omega_1, i_{X^a}i_{X^b}\omega_2]_{SN}$. In order to compare (33) and (34), we need to expand the brackets $[.,.]_{Cl}$ and $[.,.]_{SN}$ in terms of wedge products. The Clifford commutator of two differential forms can be written in terms of wedge products as follows \cite{Benn Tucker}
\begin{eqnarray}
[\omega_1,\omega_2]_{Cl}=\sum_{k=0}^{n}\frac{(-1)^{\lfloor{k/2\rfloor}}}{k!}\Bigg\{\left(\eta^k i_{X_{I(k)}}\omega_1\right)\wedge i_{X^{I(k)}}\omega_2-\left(\eta^k i_{X_{I(k)}}\omega_2\right)\wedge i_{X^{I(k)}}\omega_1\Bigg\}
\end{eqnarray}
where $\lfloor{\rfloor}$ is the floor function that takes the integer part of the argument, $\eta$ is the main automorphism of the exterior algebra that acts on a $p$-form $\alpha$ as $\eta\alpha=(-1)^p\alpha$ and $I(k)$ is a multi index. In five dimensions, higher than fifth order contractions of forms will be zero and we have the following equality for odd KY forms
\begin{equation}
[\omega_1, \omega_2]_{Cl}=2\omega_1\wedge\omega_2-i_{X_a}i_{X_b}\omega_1\wedge i_{X^a}i_{X^b}\omega_2+\frac{1}{12}i_{X_a}i_{X_b}i_{X_c}i_{X_d}\omega_1\wedge i_{X^a}i_{X^b}i_{X^c}i_{X^d}\omega_2.
\end{equation}
Similarly, other brackets in (33) can be written for odd forms in five dimensions as
\begin{equation}
[\omega_1, d\omega_2]_{Cl}=2i_{X_a}\omega_1\wedge i_{X^a}d\omega_2-\frac{1}{3}i_{X_a}i_{X_b}i_{X_c}\omega_1\wedge i_{X^a}i_{X^b}i_{X^c}d\omega_2
\end{equation}
\begin{equation}
[d\omega_1, \omega_2]_{Cl}=-2i_{X^a}\omega_2\wedge i_{X_a}d\omega_1+\frac{1}{3}i_{X^a}i_{X^b}i_{X^c}\omega_2\wedge i_{X_a}i_{X_b}i_{X_c}d\omega_1
\end{equation}
\begin{equation}
[d\omega_1, d\omega_2]_{Cl}=-2i_{X_a}d\omega_1\wedge i_{X^a}d\omega_2+\frac{1}{3}i_{X_a}i_{X_b}i_{X_c}d\omega_1\wedge i_{X^a}i_{X^b}i_{X^c}d\omega_2.
\end{equation}
The SN brackets in (34) can also be written in terms of wedge products from the definition (20) and by using KY equation (9);
\begin{equation}
[\omega_1, \omega_2]_{SN}=\frac{1}{q+1}i_{X_a}\omega_1\wedge i_{X^a}d\omega_2-\frac{1}{p+1}i_{X^a}\omega_2\wedge i_{X_a}d\omega_1.
\end{equation}
\begin{equation}
[i_{X_a}i_{X_b}\omega_1, i_{X^a}i_{X^b}\omega_2]_{SN}=\frac{1}{q+1}i_{X_a}i_{X_b}i_{X_c}\omega_1\wedge i_{X^a}i_{X^b}i_{X^c}d\omega_2-\frac{1}{p+1}i_{X^a}i_{X^b}i_{X^c}\omega_2\wedge i_{X_a}i_{X_b}i_{X_c}d\omega_1.
\end{equation}
The exterior derivative of (40) can be found as follows
\begin{eqnarray}
d[\omega_1, \omega_2]_{SN}&=&\frac{1}{q+1}di_{X_a}\omega_1\wedge i_{X^a}d\omega_2+\frac{1}{q+1}i_{X_a}\omega_1\wedge di_{X^a}d\omega_2\nonumber\\
&&-\frac{1}{p+1}di_{X^a}\omega_2\wedge i_{X_a}d\omega_1-\frac{1}{p+1}i_{X^a}\omega_2\wedge di_{X_a}d\omega_1\nonumber\\
&=&-\frac{p+q}{(p+1)(q+1)}i_{X_a}d\omega_1\wedge i_{X^a}d\omega_2+\frac{1}{q}i_{X^a}\omega_1\wedge R_{ba}\wedge i_{X^b}\omega_2-\frac{1}{p}i_{X^a}\omega_2\wedge R_{ba}\wedge i_{X^b}\omega_1\nonumber\\
&=&-\frac{p+q}{(p+1)(q+1)}i_{X_a}d\omega_1\wedge i_{X^a}d\omega_2+4\lambda^2(p+q)\omega_1\wedge\omega_2
\end{eqnarray}
where we have used (14) and curvature 2-forms in $AdS_5$ as $R_{ab}=-4\lambda^2e_a\wedge e_b$. The exterior derivative of (41) is calculated as
\begin{eqnarray}
d[i_{X_a}i_{X_b}\omega_1, i_{X^a}i_{X^b}\omega_2]_{SN}&=&di_{X_a}i_{X_b}i_{X_c}\omega_1\wedge i_{X^a}i_{X^b}i_{X^c}d\omega_2+i_{X_a}i_{X_b}i_{X_c}\omega_1\wedge di_{X^a}i_{X^b}i_{X^c}d\omega_2\nonumber\\
&&-di_{X^a}i_{X^b}i_{X^c}\omega_2\wedge i_{X_a}i_{X_b}i_{X_c}d\omega_1-i_{X^a}i_{X^b}i_{X^c}\omega_2\wedge di_{X_a}i_{X_b}i_{X_c}d\omega_1\nonumber\\
&=&-\frac{p+q-4}{(p+1)(q+1)}i_{X_a}i_{X_b}i_{X_c}d\omega_1\wedge i_{X^a}i_{X^b}i_{X^c}d\omega_2\nonumber\\
&&+24\lambda^2i_{X_a}i_{X_b}\omega_1\wedge i_{X^a}i_{X^b}\omega_2.
\end{eqnarray}
Now we have all the brackets in (33) and (34) and we can compare them for different values of $p$ and $q$. Since we consider odd forms in five dimensions, $p$ and $q$ can take values 1, 3 and 5. So, we will investigate the different cases separately as follows:

i) $p=q=1$. This case reduces to the Lie derivatives with respect to Killing vector fields because of the second part of (26) vanishes and SN bracket reduces to the Lie bracket of Killing vectors. So, the first-order symmetry operator algebra structure correspons to the following propery of Lie derivatives
\begin{equation}
[{\cal{L}}_{K_1}, {\cal{L}}_{K_2}]={\cal{L}}_{[K_1, K_2]}
\end{equation}
for two Killing vectors $K_1$ and $K_2$.

ii) $p=1$, $q=3$ or $p=3$, $q=1$ and $p=1$, $q=5$ or $p=5$, $q=1$. In those cases, since one of the KY forms is a 1-form, higher order contractions in the Clifford brackets (36)-(39) vanishes and (33) reduces to
\begin{eqnarray}
[L_{\omega_1}, L_{\omega_2}]\epsilon&=&2\lambda^2pq(\omega_1\wedge\omega_2).\epsilon+\lambda\frac{pq}{q+1}(i_{X_a}\omega_1\wedge i_{X^a}d\omega_2).\epsilon\nonumber\\
&&-\lambda\frac{pq}{p+1}(i_{X^a}\omega_2\wedge i_{X_a}d\omega_1).\epsilon-\frac{pq}{2(p+1)(q+1)}(i_{X_a}d\omega_1\wedge i_{X^a}d\omega_2).\epsilon.
\end{eqnarray}
From the same reason mentioned above, (41) and (43) also vanishes and by using (40) and (42), the operator in (34) reduces to
\begin{eqnarray}
L_{[\omega_1, \omega_2]_{KY}}\epsilon&=&\lambda\frac{pq}{q+1}(i_{X_a}\omega_1\wedge i_{X^a}d\omega_2).\epsilon-\lambda\frac{pq}{p+1}(i_{X^a}\omega_2\wedge i_{X_a}d\omega_1).\epsilon\nonumber\\
&&-\frac{pq}{2(p+1)(q+1)}(i_{X_a}d\omega_1\wedge i_{X^a}d\omega_2).\epsilon+2\lambda^2pq(\omega_1\wedge\omega_2).\epsilon.
\end{eqnarray}
By comparing (45) and (46), one can see that
\begin{equation}
[L_{\omega_1}, L_{\omega_2}]\epsilon=L_{[\omega_1, \omega_2]_{KY}}\epsilon.
\end{equation}

iii) $p=3$, $q=5$ or $p=5$, $q=3$ and $p=q=5$. Let us first choose $p=3$ and $q=5$. In this case, $\omega_2$ is the constant multiple of the volume form $z$, so we have $d\omega_2=0$. The Clifford brackets in (36)-(39) turns into
\begin{equation}
[\omega_1, \omega_2]_{Cl}=-i_{X_a}i_{X_b}\omega_1\wedge i_{X^a}i_{X^b}\omega_2
\end{equation}
\begin{equation}
[d\omega_1, \omega_2]_{Cl}=\frac{1}{3}i_{X^a}i_{X^b}i_{X^c}\omega_2\wedge i_{X_a}i_{X_b}i_{X_c}d\omega_1
\end{equation}
\begin{equation}
[\omega_1, d\omega_2]_{Cl}=[d\omega_1, d\omega_2]_{Cl}=0.
\end{equation}
However, in five dimensions with Lorentzian signature, the volume form $z$ is in the center of the Clifford algebra and it Clifford commutes with all other elements. So, we have
\begin{equation}
[\omega_1, \omega_2]_{Cl}=[d\omega_1, \omega_2]_{Cl}=0.
\end{equation}
This means that (33) is equal to zero in that case. On the other hand, (40)-(43) reduces to
\begin{equation}
[\omega_1, \omega_2]_{SN}=d[\omega_1, \omega_2]_{SN}=0
\end{equation}
\begin{eqnarray}
[i_{X_a}i_{X_b}\omega_1, i_{X^a}i_{X^b}\omega_2]_{SN}&=&-\frac{1}{p+1}i_{X^a}i_{X^b}i_{X^c}\omega_2\wedge i_{X_a}i_{X_b}i_{X_c}d\omega_1\nonumber\\
&=&-\frac{3}{p+1}[d\omega_1, \omega_2]_{Cl}\nonumber\\
&=&0
\end{eqnarray}
\begin{eqnarray}
d[i_{X_a}i_{X_b}\omega_1, i_{X^a}i_{X^b}\omega_2]_{SN}&=&24\lambda^2i_{X_a}i_{X_b}\omega_1\wedge i_{X^a}i_{X^b}\omega_2\nonumber\\
&=&-24\lambda^2[\omega_1, \omega_2]_{Cl}\nonumber\\
&=&0.
\end{eqnarray}
Those turns (34) into zero in that case, and we have the equality
\begin{equation}
[L_{\omega_1}, L_{\omega_2}]\epsilon=L_{[\omega_1, \omega_2]_{KY}}\epsilon=0.
\end{equation}
Similar reasoning gives exactly the same result for the case $p=5$ and $q=3$. For $p=q=5$, we have two volume forms and all Clifford and SN brackets will vanish, hence the result is (55) again.

iv) $p=q=3$. In that case, (36) and (39) reduce to
\begin{equation}
[\omega_1, \omega_2]_{Cl}=-i_{X_a}i_{X_b}\omega_1\wedge i_{X^a}i_{X^b}\omega_2
\end{equation}
\begin{equation}
[d\omega_1, d\omega_2]_{Cl}=\frac{1}{3}i_{X_a}i_{X_b}i_{X_c}d\omega_1\wedge i_{X^a}i_{X^b}i_{X^c}d\omega_2
\end{equation}
while (37) and (38) remain unchanged. Thus, (33) can be written as
\begin{eqnarray}
[L_{\omega_1}, L_{\omega_2}]\epsilon&=&-9\lambda^2(i_{X_a}i_{X_b}\omega_1\wedge i_{X^a}i_{X^b}\omega_2).\epsilon+\frac{9}{4}\lambda(i_{X_a}\omega_1\wedge i_{X^a}d\omega_2).\epsilon\nonumber\\
&&-\frac{3}{8}\lambda(i_{X_a}i_{X_b}i_{X_c}\omega_1\wedge i_{X^a}i_{X^b}i_{X^c}d\omega_2).\epsilon-\frac{9}{4}\lambda(i_{X^a}\omega_2\wedge i_{X_a}d\omega_1).\epsilon\nonumber\\
&&+\frac{3}{8}\lambda(i_{X^a}i_{X^b}i_{X^c}\omega_2\wedge i_{X_a}i_{X_b}i_{X_c}d\omega_1).\epsilon\nonumber\\
&&+\frac{3}{64}(i_{X_a}i_{X_b}i_{X_c}d\omega_1\wedge i_{X^a}i_{X^b}i_{X^c}d\omega_2).\epsilon.
\end{eqnarray}
The SN brackets (40), (41) and (43) are same, but since $[\omega_1, \omega_2]_{SN}$ is the volume form now, we have $d[\omega_1, \omega_2]_{SN}=0$. Hence, (34) turns into
\begin{eqnarray}
L_{[\omega_1, \omega_2]_{KY}}\epsilon&=&\frac{9}{4}\lambda(i_{X_a}\omega_1\wedge i_{X^a}d\omega_2).\epsilon-\frac{9}{4}\lambda(i_{X^a}\omega_2\wedge i_{X_a}d\omega_1).\epsilon\nonumber\\
&&-\frac{3}{8}\lambda(i_{X_a}i_{X_b}i_{X_c}\omega_1\wedge i_{X^a}i_{X^b}i_{X^c}d\omega_2).\epsilon+\frac{3}{8}\lambda(i_{X^a}i_{X^b}i_{X^c}\omega_2\wedge i_{X_a}i_{X_b}i_{X_c}d\omega_1).\epsilon\nonumber\\
&&+\frac{3}{64}(i_{X_a}i_{X_b}i_{X_c}d\omega_1\wedge i_{X^a}i_{X^b}i_{X^c}d\omega_2).\epsilon-9\lambda^2(i_{X_a}i_{X_b}\omega_1\wedge i_{X^a}i_{X^b}\omega_2).\epsilon
\end{eqnarray}
and one can see that (58) and (59) are equivalent to each other
\begin{equation}
[L_{\omega_1}, L_{\omega_2}]\epsilon=L_{[\omega_1, \omega_2]_{KY}}\epsilon.
\end{equation}

So, we exhaust all the possibilities and prove that for the KY bracket defined in (26), the first-order symmetry operators of Killing spinor equation defined in (12) constitute an algebra structure in $AdS_5$ which means that there are no higher-order symmetry operators of Killing spinor equation. Hence, we can construct all elements of the symmetry algebra and this provides a way to obtain a general solution of the Killing spinor equation by means of separation of variables. Although there are alternative methods to solve the Killing spinor equation in $AdS$ spacetimes \cite{OFarrill Gutowski Sabra}, this method gives a general approach that can also be applied to other constant curvature spacetimes with relevant modifications.

The importance of finding the solutions of the Killing spinor equation in constant curvature spacetimes relies on the fact that the maximally supersymmetric supergravity backgrounds such as Freund-Rubin backgrounds $AdS_4\times S^7$, $AdS_7\times S^4$ and $AdS_5\times S^5$ have a direct product structure in terms of them. In Freund-Rubin backgrounds, the supergravity Killing spinors which are defined as the solutions of the supergravity Killing spinor equations obtained from the variation of the gravitino field and determine the number of preserved supersymmetries in those backgrounds, can be obtained from the geometric Killing spinors (5) in component spacetimes \cite{OFarrill HackettJones Moutsopoulos Simon}. So, obtaining the solutions of the Killing spinor equation gives rise to finding supergravity Killing spinors in Freund-Rubin backgrounds.

Besides the importance of the algebra structure of first-order symmetry operators of Killing spinor equation in its own, it is also related to possible extensions of Killing superalgebras to higher-degree forms in constant curvature spacetimes. Indeed, this symmetry algebra property corresponds to one of the Jacobi identities of the extended Killing superalgebras. That gives way to the question that whether the other Jacobi identities can also be satisfied and the Killing superalgebras can be extended to include higher-degree KY forms. We will focus on this problem in the next section.

\section{Lie Superalgebras and Their Extensions}

Supersymmetric supergravity backgrounds are characterized by the number of preserved supersymmetries on them which correspond to the solutions of the supergravity Killing spinor equations. An important invariant of the supergravity backgrounds is the Lie superalgebra generated by supergravity Killing spinors. The odd part of the superalgebra corresponds to the space of Killing spinors and the even part is the Lie algebra of Killing vector fields that are constructed from supergravity Killing spinors. This Lie superalgebra is called Killing superalgebra and some geometrical properties of supergravity backgrounds can be obtained from it. So, they play an important role in the classification problem of supergravity backgrounds.

Similarly, for the manifolds that admit geometric Killing spinors, a Killing superalgebra can be constructed from Killing vector fields generated by them. However, the squaring map of Killing spinors give also higher-degree KY forms and they satisfy a graded Lie algebra structure in constant curvature spacetimes. Hence, the question of under which conditions the extension of Killing superalgebras to include KY forms is possible will be investigated in this section. Although there are some attempts to generalize the Killing superalgebras to the so-called maximal superalgebras, general constructions have not been achieved yet \cite{OFarrill HackettJones Moutsopoulos Simon, Alekseevsky Cortes Devchand Proeyen, DAuria Ferrara Lledo Varadarajan, Ferrara Porrati}. We first define Lie superalgebras and Killing superalgebras and then investigate the conditions to extend them to inlude KY forms in $AdS_5$ spacetimes in the light of the results of the previous section.

A Lie superalgebra $\mathfrak{g}=\mathfrak{g}_0\oplus\mathfrak{g}_1$ consists of an even part $\mathfrak{g}_0$ which is a Lie algebra and an odd part $\mathfrak{g}_1$ which is a $\mathfrak{g}_0$-module. A bilinear multiplication on the superalgebra is defines as
\begin{equation}
[.,.]:\mathfrak{g}_i\times\mathfrak{g}_j\longrightarrow\mathfrak{g}_{i+j}
\end{equation}
where $i, j=0,1$ mod 2. This Lie bracket satisfy the following (skew)-supersymmetry and super-Jacobi identities
\begin{eqnarray}
[a,b]&=&-(-1)^{|a||b|}[b,a]\nonumber\\
\left[a,[b,c]\right]&=&[[a, b], c]+(-1)^{|a||b|}[b, [a, c]]
\end{eqnarray}
where $a, b, c$ are elements of $\mathfrak{g}$ and $|a|$ denotes the degree of $a$ which corresponds to 0 or 1 depending on $a$ is in $\mathfrak{g}_0$ or $\mathfrak{g}_1$, respectively. For a Killing superalgebra, $\mathfrak{g}_1$ corresponds to the space of Killing spinors and $\mathfrak{g}_0$ corresponds to the Lie algebra of Killing vector fields constructed out of Killing spinors. The Lie brackets of the superalgebra are defined as follows. The Lie bracket that takes two even elements and give another even element is the ordinary Lie bracket of vector fields;
\begin{equation}
[.,.]:\mathfrak{g}_0\times\mathfrak{g}_0\longrightarrow\mathfrak{g}_0.
\end{equation}
The Lie bracket that takes one even and one odd element and gives
an odd element corresponds to the spinor Lie derivative with respect
to Killing vector fields;
\begin{equation}
{\cal{L}}:\mathfrak{g}_0\times\mathfrak{g}_1\longrightarrow\mathfrak{g}_1
\end{equation}
which is defined in (8). The Lie bracket that takes two odd elements and give an even element is the squaring map of the spinors;
\begin{equation}
(\,\,\,)_1:\mathfrak{g}_1\times\mathfrak{g}_1\longrightarrow\mathfrak{g}_0
\end{equation}
which is defined for spinors $\epsilon$ and $\kappa$ as the metric dual of the following 1-form in terms of the co-frame basis $e^a$
\begin{equation}
(\epsilon\bar{\kappa})_1=(\epsilon,e_a.\kappa)e^a
\end{equation}
and corresponds to a Killing vector, where $(\,\,\,)_1$ denotes the
projection on the 1-form component of the inhomogeneous form
$\epsilon\bar{\kappa}$ and $\bar{\kappa}$ is the dual of the spinor $\kappa$
with respect to the spinor inner product $(.,.)$. For $\epsilon=\kappa$, this
quantity is called the Dirac current $V_{\epsilon}$ of the spinor
$\epsilon$.

To obtain the consistency of the Killing superalgebra constructed above, the Jacobi identities of the Lie superalgebra must be satisfied. There are four different Jacobi identities for even and odd parts of the superalgebra. The $[\mathfrak{g}_0,\mathfrak{g}_0,\mathfrak{g}_0]$ component is the Jacobi identity of the Lie algebra of Killing vector fields and obviously satisfied. The $[\mathfrak{g}_0,\mathfrak{g}_0,\mathfrak{g}_1]$ component is the following property of the spinor Lie derivative on spinors;
\begin{equation}
[{\cal{L}}_{K_1}, {\cal{L}}_{K_2}]\epsilon={\cal{L}}_{[K_1,K_2]}\epsilon
\end{equation}
for two Killing vectors $K_1$ and $K_2$. This corresponds to the algebra structure of first-order symmetry operators of Killing spinor equation for Killing vectors in (44). The $[\mathfrak{g}_0,\mathfrak{g}_1,\mathfrak{g}_1]$ component is the compatibility of Lie derivative on differential forms and Lie derivative on spinors;
\begin{equation}
{\cal{L}}_K(\epsilon\bar{\kappa})=({\cal{L}}_K\epsilon).\bar{\kappa}+\epsilon.\overline{{\cal{L}}_K\kappa}.
\end{equation}
The $[\mathfrak{g}_1,\mathfrak{g}_1,\mathfrak{g}_1]$ component corresponds to the property that the Lie derivative of a Killing spinor with respect to its Dirac current vanishes;
\begin{equation}
{\cal{L}}_{V_{\epsilon}}\epsilon=0.
\end{equation}

To obtain an extension of the Killing superalgebra which includes higher-degree KY forms, we need to define the new brackets in the extended superalgebra and check the consistency of the Jacobi identities. We denote the extended superalgebra as $\bar{\mathfrak{g}}=\bar{\mathfrak{g}}_0\oplus\bar{\mathfrak{g}}_1$. We consider the $AdS_5$ spacetime and the brackets of KY forms and symmetry operators of Killing spinors defined on it in the previous section. So, the even part of the superalgebra consists of $\bar{\mathfrak{g}}_0=\Lambda^1V\oplus\Lambda^3V\oplus\Lambda^5V$ where $\Lambda^pV$ denotes the space of KY $p$-forms. The odd part is not changed and defined as $\bar{\mathfrak{g}}_1=S$ where $S$ is the space of Killing spinors.
The bracket of the even part is the KY bracket defined in (26) since it takes odd KY forms and gives again odd KY forms
\begin{equation}
[.,.]_{KY}:\bar{\mathfrak{g}}_0\times\bar{\mathfrak{g}}_0\longrightarrow\bar{\mathfrak{g}}_0
\end{equation}
The action of the even part to the odd part corresponds to the symmetry operators of Killing spinors defined in (12)
\begin{equation}
L:\bar{\mathfrak{g}}_0\times\bar{\mathfrak{g}}_1\longrightarrow\bar{\mathfrak{g}}_1
\end{equation}
The bracket that takes two odd elements and give an even element is the squaring map of spinors projected on odd forms
\begin{equation}
(\,\,\,)_p:\bar{\mathfrak{g}}_1\times\bar{\mathfrak{g}}_1\longrightarrow\bar{\mathfrak{g}}_0
\end{equation}
This is defined as follows. The Clifford product of a spinor $\epsilon$ and its dual $\bar{\epsilon}$ can be witten as an inhomogeneous differential form as
\begin{equation}
\epsilon\bar{\epsilon}=(\epsilon, \epsilon)+(\epsilon, e_a\epsilon)e^a+(\epsilon, e_{ba}\epsilon)e^{ab}+...+(\epsilon, e_{a_p...a_1}\epsilon)e^{a_1...a_p}+...+(-1)^{\lfloor n/2\rfloor}(\epsilon, z\epsilon)z
\end{equation}
where $( . , . )$ is the spinor innner product, $z$ is the volume form and $e^{a_1...a_p}=e^{a_1}\wedge...\wedge e^{a_p}$. This equality is called the Fierz identity and the left hand side corresponds to the spinor bilinear constructed out of $\epsilon$. The $p$-form projections on the right hand side which are defined as
\begin{equation}
(\epsilon\bar{\epsilon})_p=(\epsilon, e_{a_p...a_1}\epsilon)e^{a_1...a_p}.
\end{equation}
are called $p$-form Dirac currents. It is shown in \cite{Acik Ertem} that the $p$-form projections of the spinor bilinears of Killing spinors correspond to KY forms for $p$ odd. So, we have a well-defined bracket in (72). Now, we investigate the even-even-odd, odd-odd-odd, even-even-even and even-odd-odd Jacobi identities and find the constraints on extending the Killing superalgebra.

The $[\bar{\mathfrak{g}}_0, \bar{\mathfrak{g}}_0, \bar{\mathfrak{g}}_1]$ Jacobi identity corresponds to the algebra of first-order symmetry operators of Killing spinor equation which was the main motivation for investigating the extension of Killing superalgebras in $AdS_5$. It is proved in the previous section that we have the equality
\begin{equation}
[L_{\omega_1}, L_{\omega_2}]\epsilon=L_{[\omega_1, \omega_2]_{KY}}\epsilon
\end{equation}
for two odd KY forms $\omega_1$ and $\omega_2$. $[\bar{\mathfrak{g}}_0, \bar{\mathfrak{g}}_0, \bar{\mathfrak{g}}_0]$, $[\bar{\mathfrak{g}}_0, \bar{\mathfrak{g}}_1, \bar{\mathfrak{g}}_1]$ and $[\bar{\mathfrak{g}}_1, \bar{\mathfrak{g}}_1, \bar{\mathfrak{g}}_1]$ Jacobi identities are written as follows
\begin{equation}
[\omega_1, [\omega_2, \omega_3]_{KY}]_{KY}+[\omega_2, [\omega_3, \omega_1]_{KY}]_{KY}+[\omega_3, [\omega_1, \omega_2]_{KY}]_{KY}=0.
\end{equation}
\begin{equation}
[\omega, (\epsilon\bar{\kappa})]_{KY}=L_{\omega}\epsilon.\bar{\kappa}+\epsilon.\overline{L_{\omega}\kappa}.
\end{equation}
\begin{equation}
\left(L_{(\epsilon\bar{\epsilon})_1}+L_{(\epsilon\bar{\epsilon})_3}+L_{(\epsilon\bar{\epsilon})_5}\right)\epsilon=0.
\end{equation}
However, these identities imply that KY 3-forms and 5-form have only vanishing brackets with other elements of the superalgebra. This means that the Killing superalgebra of $AdS_5$ cannot be extended to include higher-degree forms since the brackets defined in (70), (71) and (72) give non-zero values for those forms. Since we can define the superalgebra brackets (70)-(72) consistently, the Killing spinors and odd KY forms constitute a superalgebra structure with respect to these brackets, but it does not correspond to a Lie superalgebra because of the reason that all Jacobi identities are not satisfied. This implies the existence of an extended rigid supersymmetry algebra rather than an extended supergravity Lie superalgebra in $AdS_5$ background.

\section{Conclusion}

First-order symmetry operators of Killing spinor equation are constructed from odd KY forms. In constrast to the symmetric Killing tensors, KY forms do not have a Lie algebra structure in general. However, they satisfy a graded Lie algebra in constant curvature spacetimes with respect to the SN bracket. Moreover, odd KY forms themselves constitute a Lie algebra under this bracket. Yet this does not mean that the first-order symmetry operators of Killing spinors can satisfy a closed algebra with respect to the SN bracket. We show that a symmetry algebra of first-order symmetry operators of Killing spinors can be constructed in five dimensional constant curvature spatimes, especially in $AdS_5$, by modifying the SN bracket to the KY bracket defined in (26). This procedure is also relevant for other five or lower dimensional constant curvature spacetimes. Finding a first-order symmetry operator algebra means that the solutions of the Killing spinor equation can be investigated by using symmetry properties. Since Killing spinors can be used to construct supergravity Killing spinors, which are parallel spinors with respect to a supergravity spin connection, by the method of cone construction, this will also lead to the solutions of supergravity Killing spinors. Indeed, the method of finding the symmetry operator algebra of Killing spinors can be applied to other higher dimensional spacetimes by taking different modifications of SN bracket with relevant coefficients.

The equation that defines the symmetry operator algebra property of Killing spinors corrresponds to a Jacobi identity in extended Killing superalgebras. Killing superalgebras contain Killing spinors and Killing vector fields constructed out of them by the squaring map. However, the squaring map of Killing spinors also generates higher-degree KY forms and the inclusion of these higher-degree forms to the Killing superalgebra can be possible. We showed that in $AdS_5$ backgrounds, the space of Killing spinors and odd KY forms constitute a superalgebra but not a Lie superalgebra since the Jacobi identities of extended Killing superalgebra constrain the Lie brackets defined for higher-degree forms and the Killing superalgebra cannot be extended in this case. The constructed rigid superalgebra may have relations to a supersymmetric field theory in the background rather than a supergravity theory.

\begin{acknowledgments}

The author thanks Jose M. Figueroa-O`Farrill, Andrea Santi and \"{O}zg\"{u}r A\c{c}{\i}k for inspiring discussions on various subjects. He also thanks School of Mathematics of The University of Edinburgh for the kind hospitality and providing a fruitful scientific atmosphere during his stay in Edinburgh. This work is supported by the Scientific and Technological Research Council of Turkey (T\"{U}B\.{I}TAK) grant B\.{I}DEB 2219.

\end{acknowledgments}

%\references%

\end{document}